# Role of friction in pattern formation in oscillated granular layers

Sung Joon Moon,* J. B. Swift, and Harry L. Swinney

*Center for Nonlinear Dynamics and Department of Physics, University of Texas, Austin, Texas 78712, USA*

Particles in granular flows are often modeled as frictionless (smooth) inelastic spheres; however, there exist no frictionless grains, just as there are no elastic grains. Our molecular dynamics simulations reveal that friction is essential for realistic modeling of vertically oscillated granular layers: simulations of frictionless particles yield patterns with an onset at a container acceleration about 30% smaller than that observed in experiments and simulations with friction. More importantly, even though square and hexagonal patterns form for a wide range of the oscillation parameters in experiments and in our simulations of frictional inelastic particles, only stripe patterns form in the simulations without friction, even if the inelasticity is increased to obtain as much dissipation as in frictional particles. We also consider the effect of particle friction on the shock wave that forms each time the granular layer strikes the container. While a shock wave still forms for frictionless particles, the spatial and temporal dependence of the hydrodynamic fields differ for the cases with and without friction.

## I. INTRODUCTION

Collisional dissipation between grains in rapid granular flows arises from both inelasticity and surface friction. Hydrodynamic models for frictionless inelastic hard disks and spheres have been proposed [1–4] as simple models for granular flows, and the agreement between such equations proposed by Jenkins and Richman [3] and molecular dynamics (MD) simulations of frictionless inelastic hard spheres has been found to be good even when a shock wave is present [5,6]. Hydrodynamic models for slightly frictional inelastic hard disks and spheres also have been proposed [7–9]; however, the extent to which friction has to be considered is not known.

The role of friction is difficult to distinguish from that of inelasticity in experiments since there are no frictionless grains; however, MD simulations can be conducted with or without friction, and such simulations have revealed that, even in the dilute limit, the behavior of a collection of frictional inelastic particles differs from that of frictionless particles: In vibrofluidized disks and spheres, the scaling exponent of the mean height as a function of the characteristic velocity [10] and the single-particle distribution functions [11] exhibit qualitative differences depending on the presence of friction. Some recent numerical studies [12–14] have referred to dilute frictionless inelastic hard spheres as granular gases; while frictionless models may yield insights into some granular flow phenomena, there is no way of knowing *a priori* how relevant the predictions of such models are for real (frictional) granular gases.

We use here a previously validated MD simulation [15,16] to study the role of friction in patterns and shock waves formed in vertically oscillated granular layers. We study how the friction affects the formation and stability of spatial patterns. We also examine the assumption that frictional effects may be accounted for by adjusting the inelasticity in some granular flows. The rest of the paper is organized as follows. Section II describes our system, and Sec. III describes the collision model implemented in our MD simulation. Results from our simulations are presented in Sec. IV and discussed in Sec. V.

## II. SYSTEM AND METHOD

We simulate shallow granular layers (with a layer depth less than ten-particle diameters) in an evacuated container, subject to a sinusoidal oscillation in the direction of gravity, with an amplitude $A$ and frequency $f$. Various subharmonic standing wave patterns have been observed depending on the nondimensional acceleration amplitude $\Gamma = A(2\pi f)^2/g$ and the nondimensional oscillation frequency $f^* = f\sqrt{H/g}$ [17], where $g$ is the acceleration due to gravity, and $H$ is the depth of the layer at rest. Our MD simulation models grains as frictional (or frictionless) inelastic hard spheres, and it implements the collision model of Walton [18].

We simulate layers of three different geometries. First, in our simulations examining the onset and stability of the patterns (Secs. IV A and IV B), we consider 8900 grains in a rectangular box of area $200\sigma \times 10\sigma$, where $\sigma$ is the diameter of the grain ($H \approx 4\sigma$). Periodic boundary conditions are used in both horizontal directions. We call this a quasi-two-dimensional (quasi-2D) layer, where one of the horizontal directions ($y$ direction) is too short to form a pattern. In the longer direction ($x$ direction) the wavelength of the pattern is about $20\sigma$ at $f^* = 0.3$, which is the frequency we use for this case. Second, we consider layers in square boxes of large aspect ratio ($L/H \gg 1$, where $L$ is the horizontal size of the container) in both horizontal directions. The square box geometry is fully 3D and computationally far less efficient than the quasi-2D geometry, but with the fully 3D geometry we obtain the various 2D spatial patterns observed in experiments. Our square boxes have bottom area $200\sigma \times 200\sigma$ with 181 390 particles ($H \approx 4\sigma$) and $100\sigma \times 100\sigma$ with 65 500 particles ($H \approx 6\sigma$). Again periodic boundary conditions are used in both horizontal directions (Secs. IV B and IV C). Finally, in our study of shock waves (Sec. IV D) we use a layer of small horizontal size (3938 particles in a box of $20\sigma \times 20\sigma$ bottom area with horizontal periodic boundary conditions; $H \approx 10\sigma$). The small horizontal dimensions are chosen so that the layer does not form a pattern; this is the same system that was used in Ref. [5].

---

*Electronic address: moon@chaos.utexas.edu

## III. COLLISION MODEL

Walton [18] simplified the collision model originally proposed by Maw *et al.* [19]. In Walton's model, the postcollisional translational and angular velocities depend on the three parameters:

(1) The normal coefficient of restitution $e$ ($0 \leq e \leq 1$).

(2) The coefficient of sliding friction $\mu$, which relates the tangential force to the normal force at collision using Coulomb's law, and then determines the tangential coefficient of restitution $\beta$ ($-1 \leq \beta \leq 1$).

(3) The maximum tangential coefficient of restitution $\beta_0$, which represents the tangential restitution of the surface velocity for a rolling contact. We vary the above parameters for grain-grain collisions (denoted by superscript $g$) and grain-wall collisions (superscript $w$) independently. By grain-wall collisions we mean collisions of grains with the container bottom, as there are no sidewalls in any of our simulations.

At collision, it is convenient to decompose the relative colliding velocities into the components normal ($\mathbf{v}_n$) and tangential ($\mathbf{v}_t$) to the relative displacement vector $\hat{\mathbf{r}}_{12} \equiv (\mathbf{r}_1 - \mathbf{r}_2)/|\mathbf{r}_1 - \mathbf{r}_2|$, where $\mathbf{r}_1$ and $\mathbf{r}_2$ are the position vectors of grains 1 and 2, respectively, and a corresponding notation is used for the velocity vector $\mathbf{v}$:

$$\mathbf{v}_n = (\mathbf{v}_{12} \cdot \hat{\mathbf{r}}_{12})\hat{\mathbf{r}}_{12} \equiv v_n \hat{\mathbf{r}}_{12}, \quad (1)$$

$$\mathbf{v}_t = \hat{\mathbf{r}}_{12} \times (\mathbf{v}_{12} \times \hat{\mathbf{r}}_{12}) = \mathbf{v}_{12} - \mathbf{v}_n. \quad (2)$$

The relative surface velocity at collision $\mathbf{v}_s$ for monodisperse spheres of diameter $\sigma$ is

$$\mathbf{v}_s = \mathbf{v}_t + \frac{\sigma}{2}\hat{\mathbf{r}}_{12} \times (\mathbf{w}_1 + \mathbf{w}_2) \equiv v_s \hat{\mathbf{v}}_s, \quad (3)$$

where $\mathbf{w}_1$ and $\mathbf{w}_2$ are the angular velocities of the grains 1 and 2, respectively.

For monodisperse spheres of unit mass and diameter $\sigma$, linear and angular momenta conservation and the definitions of the normal coefficient of restitution $e \equiv -v_n^*/v_n$ and the tangential coefficient of restitution $\beta \equiv -v_s^*/v_s$ (where superscript * indicates postcollisional velocities, and no superscript is used for precollisional velocities) give the changes in the velocities at collision:

$$\Delta \mathbf{v}_{1n} = -\Delta \mathbf{v}_{2n} = \frac{1}{2}(1+e)\mathbf{v}_n, \quad (4)$$

$$\Delta \mathbf{v}_{1t} = -\Delta \mathbf{v}_{2t} = \frac{K(1+\beta)}{2(K+1)}\mathbf{v}_s, \quad (5)$$

$$\Delta \mathbf{w}_1 = -\Delta \mathbf{w}_2 = \frac{(1+\beta)}{\sigma(K+1)}\hat{\mathbf{r}}_{12} \times \mathbf{v}_s, \quad (6)$$

where $K = 4I/\sigma^2$ is a geometrical factor that relates the momentum transfer from the translational to the rotational degrees of freedom, and $I$ is the moment of inertia about the center of the grain. For a sphere of uniform density, $K$ is 2/5.

We use a velocity-dependent coefficient of restitution to account for the viscoelasticity of the real grains:

$$e = \max\left[e_0, 1 - (1-e_0)\left(\frac{v_n}{\sqrt{g\sigma}}\right)^{3/4}\right], \quad (7)$$

where $e_0$ is a constant value, a fitting parameter determined from the comparison with the experiment. An accurate functional form for $e$ is not known. The form in Eq. (7) has been previously found to reproduce observed granular patterns [15]. Our results are insensitive to the functional form for $e$, as long as $e$ approaches unity for vanishing colliding velocity; this behavior helps avoid successive collisions within a time interval comparable to machine precision, which stops the simulation. The convergence to zero time between collisions for inelastic hard sphere models with a velocity independent $e$ is a phenomenon known as inelastic collapse [20].

In collisions between real grains, not only is the relative surface velocity reduced, but also the stored tangential strain energy in the contact region can often reverse the direction of the relative surface velocity. To account for this effect, $\beta$ could be positive, leading to the range of $\beta$ as $[-1,1]$. Furthermore, there are two kinds of frictional interaction at collisions, sliding and rolling contact. The following formula for $\beta$ includes the above essential behaviors of the friction:

$$\beta = \min\left[\beta_0, -1 + \mu(1+e)\left(1+\frac{1}{K}\right)\frac{v_n}{v_s}\right]. \quad (8)$$

For sliding friction, the tangential impulse is assumed to follow Coulomb's friction law: the normal impulse multiplied by $\mu$. When $\beta$ is identically negative unity ($\beta_0 = -1$ and $\mu = 0$, or simply $\mu = 0$), it models the frictionless interaction. For the special case $v_s = 0$, the collision is treated as frictionless.

This collision model is still a simplification of the real interaction; however, this model has been found to be accurate enough to reproduce many phenomena; with the values $e_0^g = e_0^w = 0.7$, $\beta_0^g = \beta_0^w = 0.35$, and $\mu^g = \mu^w = 0.5$, this model has quantitatively reproduced the patterns in oscillated layers of 165 $\mu$m lead spheres for a wide range of control parameters [15,16].

## IV. RESULTS

As the normal coefficient of restitution ($e_0^g$ or $e_0^w$) is varied in simulations of oscillated layers of frictional inelastic hard spheres, transitions between different patterns occur at slightly different values of control parameters, and the characteristics of patterns (for instance, the amplitude and wavelength) change as well; however, the overall topology of the phase diagram remains the same, as long as the inelasticity is not too small or the layer is not too shallow. In this section, we vary the inelasticity and friction parameters for grain-grain collisions and grain-wall collisions independently to study how friction affects the pattern formation and stability. We also study how the friction changes the characteristics of the shock wave that forms at each impact.

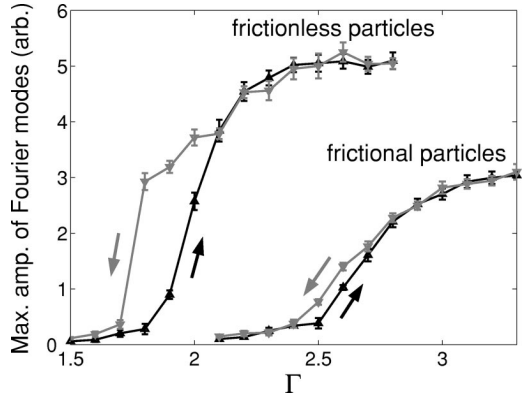

FIG. 1. The increase in amplitude of the most intense Fourier mode of the height of the layer (see text), as a function of the control parameter $\Gamma$, shows the emergence of a pattern in simulations of a quasi-2D layer of frictional inelastic particles ($e_0^g = e_0^w = 0.7, \beta_0^g = \beta_0^w = 0.35, \mu^g = \mu^w = 0.5$) and of frictionless particles ($e_0^g = e_0^w = 0.7, \beta_0^g = \beta_0^w = -1, \mu^g = \mu^w = 0$). Black lines are for increasing $\Gamma$, and gray lines are for decreasing $\Gamma$, while $f^*$ is fixed at 0.3. A pattern emerges at smaller $\Gamma$ and the pattern amplitude is larger in the frictionless case. The transition is weakly hysteretic in both cases. For each $\Gamma$, the amplitude was averaged over ten $f/2$-strobed cycles after the amplitude was saturated, and error bars indicate the standard deviation.

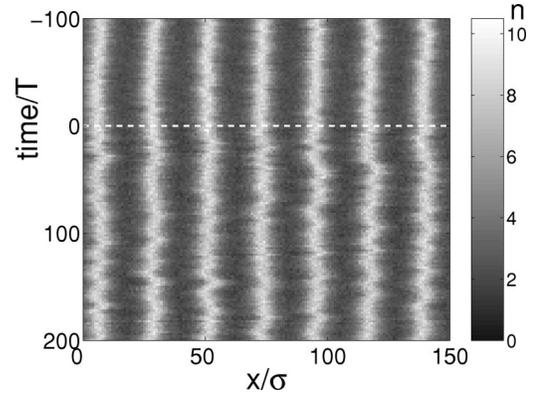

FIG. 2. Evolution of a pattern in a quasi-2D layer illustrates that the peaks of the pattern fluctuate around the average locations more in the frictionless model than in the frictional model. The pattern, strobed at $f/2$, is shown for 300 periods $T$ of the container oscillation; only part of the layer is shown. The friction was suddenly removed from the simulation at $t=0$ (indicated by a white dashed line); ($e_0^g = e_0^w, \beta_0^g = \beta_0^w, \mu^g = \mu^w$) were changed from (0.7,0.35,0.5) to (0.5,-1,0), while the control parameters remained the same at $\Gamma = 3.0$ and $f^* = 0.3$. Both $e_0^g$ and $e_0^w$ were reduced to keep the amplitude and wavelength of the pattern nearly the same. The gray scale indicates the vertically integrated number of particles averaged over the $y$ direction.

### A. Primary onset and hysteresis

The primary onset of the pattern, the transition from a featureless flat state to an $f/2$ square or stripe subharmonic pattern, occurs at a critical acceleration $\Gamma_c \approx 2.5$ [17]. This transition is weakly hysteretic, and the value of $\Gamma_c$ varies within 10% in experiments on layers of different materials or different depths [21].

We obtain the amplitude of the pattern as a function of the control parameter $\Gamma$ around the onset of patterns from simulations of two kinds of quasi-2D layers, consisting only of frictional ($\mu^g = \mu^w = 0.5$) and only of frictionless ($\mu^g = \mu^w = 0$) inelastic hard spheres (Fig. 1). We represent the pattern's amplitude by the amplitude of the most intense Fourier mode of the envelope of the layer, where the envelope is defined as the height where the volume fraction drops down to 0.2, when the container is at its equilibrium position and is moving upward. For each $\Gamma$, we wait for 40 cycles for the amplitude to saturate, and then average it over the next ten $f/2$-strobed frames. We first slowly decrease $\Gamma$ from 3.5 (2.8) down to 2.1 (1.5) for a layer of frictional (frictionless) spheres, in steps of 0.1, and then increase $\Gamma$ back to 3.5 (2.8), while $f^*$ remains at 0.3.

For both layers there is a hysteretic transition to patterns, as Fig. 1 illustrates. The amplitude of the most intense Fourier mode below the onset is not infinitesimally small compared to that above the onset, as granular fluids exhibit strong fluctuations around average quantities due to relatively small number of particles [22,23]. The transition is hysteretic in both cases, and $\Gamma_c$ decreases by about 30% in the absence of friction. At comparable values of the reduced control parameter $(\Gamma - \Gamma_c)/\Gamma_c$, the pattern's amplitude in the frictionless case is nearly twice larger than that in the frictional case. In layers of frictionless particles, the amplitude and wavelength fluctuate more than those in the frictional case (see Fig. 1); the pattern is apparently less stable in the frictionless case.

### B. Stability of the pattern

We first use the same quasi-2D layer as in the preceding section to examine how the absence of friction affects the stability of the pattern. When the grain-grain and grain-wall friction are removed from the simulation, peaks of the pattern fluctuate around the average locations more strongly than in the frictional case (Fig. 2); however, there are no qualitative changes of the pattern, as there exists only one type of pattern in this geometry. A pattern in a quasi-2D layer is *robust* due to geometrical constraints.

The fully 3D motion of the particles in square boxes leads to different 2D patterns, depending on $\Gamma$ and $f^*$. We focus here on the stability of square patterns, but we observe similar results for hexagonal patterns. We first obtained a stable $f/2$ square pattern using the same material coefficients (inelasticity and friction parameters) as in Ref. [15], and then we used that pattern as the initial condition for the different cases in Fig. 3. In Fig. 3(b), only the grain-wall friction was removed, while other properties were kept the same; in about 20 cycles, the pattern lost its stability and disorganized into randomly moving peaks. The pattern lost long range order in such a way that a shear mode became resonantly excited and the lattice melted, when the peaks of the pattern are considered as the lattice points; this melting phenomenon was the subject of Ref. [24]. We find that the square pattern is stable only for $\mu^w>0.3$ in this case.

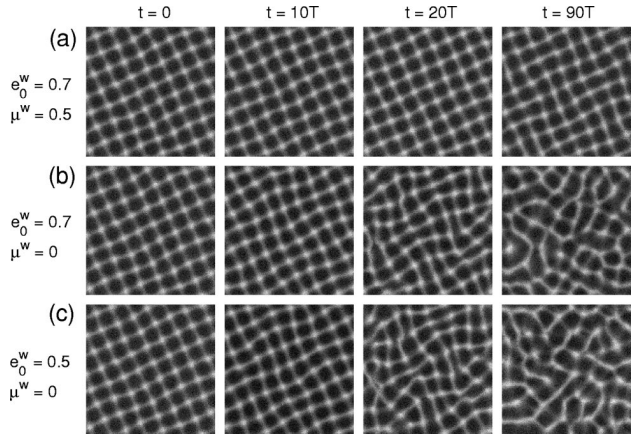

FIG. 3. A square pattern loses stability when the grain-wall friction is removed ($\mu^w=0$), even if the dissipation is increased by decreasing $e_0^w$. A stable $f/2$ square pattern was first obtained from a simulation of 181 390 frictional particles ($e_0^g=e_0^w=0.7$, $\beta_0^g=\beta_0^w=0.35$, and $\mu^g=\mu^w=0.5$) in a square box of each side $200\sigma$ with periodic boundary conditions ($\Gamma=3.0, f^*=0.27$). This square pattern was then used as the initial condition for the following three cases: (a) A further continuation of the simulation; (b) $\mu^w$ was reduced to 0 at $t=0$ while other parameters remained the same; (c) $\mu^w$ and $e_0^w$ were reduced to 0 and 0.5 respectively at $t=0$. The pattern became disordered both in (b) and (c). The gray scale indicates the vertically integrated number of particles at each location; white regions correspond to the peaks of the square patterns.

To examine the possibility that properly adjusted inelasticity could take over the effect of friction, the grain-wall coefficient of restitution $e_0^w$ was reduced to 0.5 [Fig. 3(c)]. The observed development of disorder demonstrates that the square pattern cannot be stabilized simply by reducing $e_0^w$. Even reducing $e_0^w$ to 0.1 did not stabilize the pattern. Friction plays an independent role from that of the inelasticity. While only grain-wall friction was removed in Figs. 3(b) and 3(c), we observed the same qualitative behavior when also the grain-grain friction was removed; the square pattern disorganized in a similar way as in Figs. 3(b) and 3(c).

We repeated the above numerical experiments by imposing both frictional ($e_0^w=0.7$, $\beta_0^w=0.35$, and $\mu^w=0.5$) and frictionless ($e_0^w=0.5$ and $\mu^w=0$) rigid sidewalls, rather than periodic boundaries. We did not observe any qualitative differences; the sidewalls did not stabilize the pattern, which confirmed that the above development of disorder arises from the absence of friction.

### C. 2D patterns in the absence of any friction

In the preceding section we considered the effect of removing the grain-wall friction while retaining the friction between grains. Now we consider particles with no friction at all. Starting with conditions that would yield square, stripe, or hexagonal patterns for frictional particles, we observe the evolution for layers of frictionless particles (Fig. 4). The conditions that yield squares for particles with friction evolve for frictionless particles into stripelike structures [Fig. 4(a)]; in larger containers this pattern becomes similar to the ones in Figs. 3(b) and 3(c). Starting with conditions that yield stripes for frictional patterns, we find that the frictionless particles also form stripes, but the stripes take longer to emerge than with friction [Fig. 4(b)]. The stripes formed by particles with friction are stationary, but the stripes formed by frictionless particles oscillate perpendicularly to their equilibrium positions; the oscillation period for the conditions in Fig. 4(b) is about $40T$, where $T=1/f$. Finally, starting with conditions that yield hexagonal patterns for frictional particles, we find that the frictionless particles evolve instead to a pattern of randomly moving peaks that occasionally look like stripes [Fig. 4(c)].

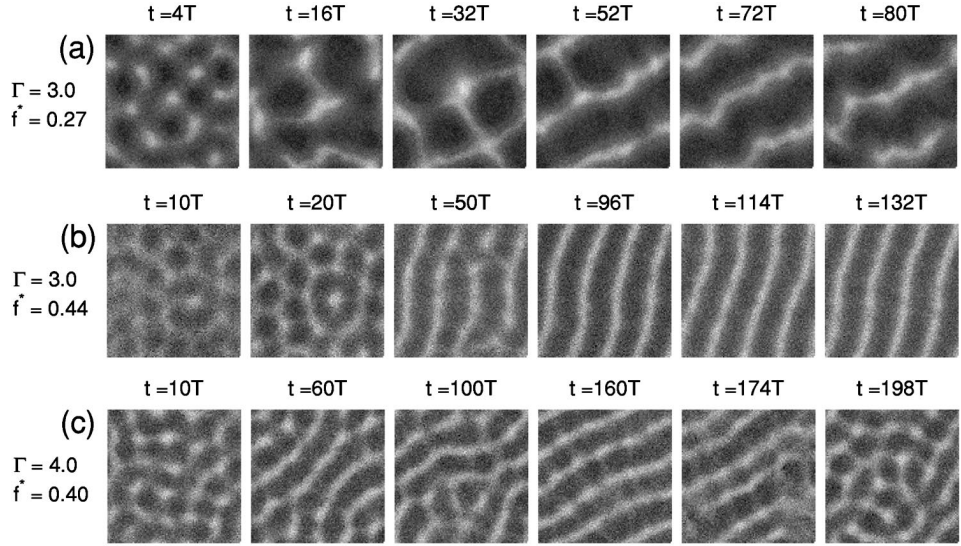

FIG. 4. Simulations of 65 500 frictionless particles ($e_0^g=e_0^w=0.7$, $\beta_0^g=\beta_0^w=-1$, and $\mu^g=\mu^w=0$) in a square box of sides $100\sigma$ illustrate that a stripe pattern is the only stable pattern that forms for frictionless particles. In each case the container was initially oscillating at conditions that yielded a flat patternless layer. Then at $t=0$, $\Gamma$ and $f^*$ were switched to values that for frictional particles would yield within 20 cycles (a) a square pattern, (b) a stripe pattern, and (c) a hexagonal pattern, each oscillating at $f/2$ [15]. Instead, for frictionless particles we observed that (a) an unstable stripelike structure slowly forms, (b) a stripe pattern slowly forms with peaks whose positions oscillate, and (c) peaks move around irregularly, sometimes forming a stripelike pattern ($t=160T$) and sometimes a cellular pattern ($t=198T$).

To see if increased dissipation, achieved by reducing the normal coefficient of restitution, could stabilize patterns for the frictionless particles, we reduced the values for $e_0^g$ and $e_0^w$ down to 0.3, but this did not yield stable patterns; hence inelasticity cannot substitute for the role of friction in the pattern formation, either. To determine how friction affects the overall dissipation, we have measured the collision rates for different parameter values. For the control parameters in Figs. 4(a) and 4(b) with friction ($e_0^g=e_0^w=0.7$, $\beta_0^w=0.35$, and $\mu^g=\mu^w=0.5$), the average number of collisions per grain during a cycle was 204 and 89, respectively, and the ratio of the average rotational kinetic energy per particle to its translational counterpart during a cycle was 0.029 and 0.044. When frictionless spheres are simulated for the control parameters in Figs. 4(a) and 4(b) ($e_0^g=e_0^w=0.7$ and $\mu^g=\mu^w=0$), the collision rates decrease by a factor of 2, and become 99 and 44, respectively. Thus, friction nearly doubles the collision rate; it increases the overall dissipation significantly. When the normal coefficients of restitution (both $e_0^g$ and $e_0^w$) were reduced to 0.5 for the frictionless cases ($\mu^g=\mu^w=0$), the number of collisions (203 and 100) became comparable with the frictional case ($e_0^g=e_0^w=0.7$ and $\mu^g=\mu^w=0.5$) of the control parameters of Figs. 4(a) and 4(b); however, the layer still did not form a pattern.

### D. Shock wave propagation

A shock wave is formed each time the granular layer strikes the plate, and the shock propagates up through the layer and decays as it propagates. Bougie *et al.* [5] studied these shock waves in MD simulations of frictionless inelastic hard spheres and in numerical simulations of continuum equations that did not include terms arising from friction; the simulations by the two approaches were in good agreement. We now examine how friction changes this shock wave. These simulations are conducted for a box with sufficiently small horizontal dimensions so that patterns do not form. Thus, following Bougie *et al.* [5], we can average the granular temperature $T_g$ and volume fraction $\nu$ in the horizontal directions and consider only their dependence on the distance $z$ above the lowest height of the oscillating plate.

We find that the $z$ dependence of the temperature and volume fraction for particles with friction is considerably different from that for frictionless particles (Fig. 5). For particles with friction, the particles undergo more collisions, the layer gets more compact, and the layer strikes the plate at a later time during a cycle. However, the shock wave formation itself and overall qualitative features of the shock are the same for frictional and frictionless particles.

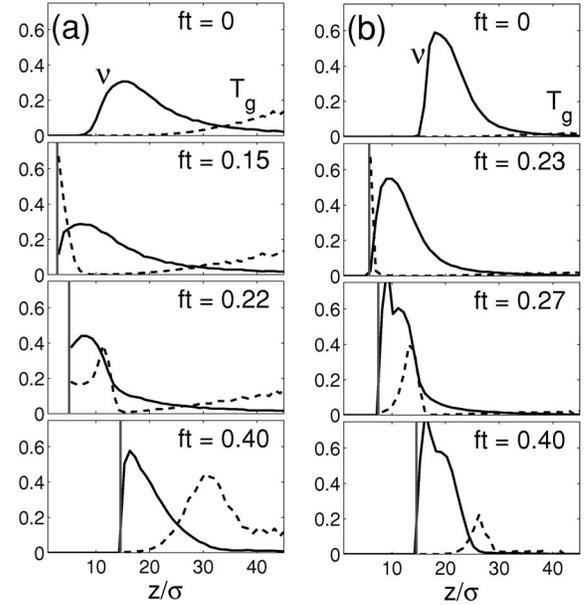

FIG. 5. A shock wave forms each time the granular layer strikes the plate, which begins to happen at $ft\approx0.15$ ($ft\approx0.23$) for the frictionless (frictional) particles, where $ft=0$ corresponds to the time when the plate is at its lowest height. The granular volume fraction $\nu$ (solid lines) and granular temperature $T_g$ (dashed lines; units of $g\sigma/50$) for layers of (a) frictionless spheres ($\beta_0^g=\beta_0^w=-1$ and $\mu^g=\mu^w=0$) and (b) frictional particles ($\beta_0^g=\beta_0^w=0.35$ and $\mu^g=\mu^w=0.5$). The times in the second and third pairs of panels are chosen to show the layer at about the same position relative to the plate, illustrating that the frictional layer collides with the plate at a later time than the layer of frictionless particles. The conditions are the same as in Ref. [5]: $e_0^g=e_0^w=0.9, \Gamma=3.0, f^*=0.3$, and $H=10\sigma$ in a box of horizontal dimensions $20\sigma\times20\sigma$.

### V. CONCLUSIONS

We conclude that friction should not be neglected for realistic modeling of pattern formation, even though inelasticity is certainly the most distinguishing characteristic of granular materials, and frictionless inelastic hard spheres provide an important simple model for granular materials. Some phenomena observed in experiments and in simulations of frictional particles, including the parametric sloshing motion of particles and the formation of shock waves, are still found in simulations without friction, but the details of these phenomena for frictionless particles are significantly different from the properties for particles with friction. We find that only stripe patterns are stable in oscillated granular layers without friction (Fig. 4).

We have found that increased inelasticity cannot substitute for the effect of friction [Fig. 3(c)]; friction is not merely an additional mechanism of dissipation. Even a small amount of friction increases the overall dissipation significantly not because the fraction of frictional dissipation is significant in each collision, but because the friction reduces the grain mobility and increases the overall collision rate (see Sec. IV C); for the cases in this paper, the average rotational kinetic energy for frictional particles is typically 30 times smaller than the translational kinetic energy. With friction, the direction of a postcollisional trajectory depends on the relative surface velocity at collision as well as the surface property, while there is only one possible direction in the frictionless case. This dependence of the transmission of force or of kinetic energy on friction is apparently critical for the formation of square and hexagonal patterns (Fig. 3), but we do not know exactly how this mechanism stabilizes those patterns.

The role of friction has not been systematically studied in experiments on oscillated granular layers, but it has been observed that the pattern formation can depend on the surface properties of grains. In experiments with grains cleaned with acid or contaminated by dust or added powder, the value of $\Gamma_c$ marking the onset of patterns changed, even in the absence of static charge effects [21]. We propose that this happened because friction depends on surface preparation.

## ACKNOWLEDGMENTS

The authors thank Jon Bougie, Daniel Goldman, and W. D. McCormick for helpful discussions. This work was supported by U.S. DOE Grant No. DE-FG-0393ER14312 and Texas Advanced Research Program Grant No. ARP-055-2001.